\documentclass[nohyperref]{article}

\usepackage{microtype}
\usepackage{graphicx}
\usepackage{subfigure}
\usepackage{booktabs} % for professional tables
\usepackage{pgfplots}
\pgfplotsset{compat=1.16}
\usepackage{tikz}
\usepackage{hyperref}

\usepackage[accepted]{icml2022}

\usepackage{amsmath}
\usepackage{amssymb}
\usepackage{mathtools}
\usepackage{amsthm}

\usepackage[capitalize,noabbrev]{cleveref}

\theoremstyle{plain}

\theoremstyle{definition}

\theoremstyle{remark}

\usepackage[textsize=tiny]{todonotes}

\icmltitlerunning{Multitask vocal burst modeling with ResNets and pre-trained paralinguistic Conformers}

\begin{document}

\twocolumn[
\icmltitle{Multitask vocal burst modeling with ResNets and pre-trained paralinguistic Conformers}

% It is OKAY to include author information, even for blind
% submissions: the style file will automatically remove it for you
% unless you've provided the [accepted] option to the icml2022
% package.

% List of affiliations: The first argument should be a (short)
% identifier you will use later to specify author affiliations
% Academic affiliations should list Department, University, City, Region, Country
% Industry affiliations should list Company, City, Region, Country

% You can specify symbols, otherwise they are numbered in order.
% Ideally, you should not use this facility. Affiliations will be numbered
% in order of appearance and this is the preferred way.
\icmlsetsymbol{equal}{*}

\begin{icmlauthorlist}
\icmlauthor{Josh Belanich}{comp}
\icmlauthor{Krishna Somandepalli}{comp}
\icmlauthor{Brian Eoff}{comp}
\icmlauthor{Brendan Jou}{comp}
\end{icmlauthorlist}

\icmlaffiliation{comp}{Google Research}

\icmlcorrespondingauthor{Josh Belanich}{joshbelanich@google.com}
%\icmlcorrespondingauthor{Firstname2 Lastname2}{first2.last2@www.uk}

% You may provide any keywords that you
% find helpful for describing your paper; these are used to populate
% the "keywords" metadata in the PDF but will not be shown in the document
\icmlkeywords{Machine Learning, ICML}

\vskip 0.3in
]

% this must go after the closing bracket ] following \twocolumn[ ...

% This command actually creates the footnote in the first column
% listing the affiliations and the copyright notice.
% The command takes one argument, which is text to display at the start of the footnote.
% The \icmlEqualContribution command is standard text for equal contribution.
% Remove it (just {}) if you do not need this facility.

\printAffiliationsAndNotice{}  % leave blank if no need to mention equal contribution
% \printAffiliationsAndNotice{\icmlEqualContribution} % otherwise use the standard text.

\begin{abstract}
This technical report presents the modeling approaches used in our submission to the ICML Expressive Vocalizations Workshop \& Competition multitask track (\textsc{ExVo-MultiTask}). We first applied image classification models of various sizes on mel-spectrogram representations of the vocal bursts, as is standard in sound event detection literature. Results from these models show an increase of 21.24\% over the baseline system with respect to the harmonic mean of the task metrics, and comprise our team's main submission to the \textsc{MultiTask} track. We then sought to characterize the headroom in the \textsc{MultiTask} track by applying a large pre-trained Conformer model that previously achieved state-of-the-art results on paralinguistic tasks like speech emotion recognition and mask detection. We additionally investigated the relationship between the sub-tasks of emotional expression, country of origin, and age prediction, and discovered that the best performing models are trained as single-task models, questioning whether the problem truly benefits from a multitask setting.
\end{abstract}

\section{Introduction}

Vocal bursts are an important but often overlooked component of paralanguage in affective computing applications. They are ubiquitous in human communication, and have been studied as channels for emotional expression \cite{hawk2009worth, cowen2019mapping}. Due to their prevalence in natural speech, predicting emotion perception of vocal bursts is an important complement to in-the-wild speech emotion recognition. The Hume Vocal Bursts dataset (\textsc{HUME-VB}) is a recent dataset for studying these paralinguistic phenomena covering four globally diverse regions across 3 languages: USA (English), China (Mandarin), South Africa (English), and Venezuela (Spanish).

We explore the suitability of ResNet \cite{he2016deep} image classifiers of various sizes in predicting the perceived emotional expression of vocal bursts and subject attributes in \textsc{HUME-VB}. We turn to ResNet architectures given their competitive history in sound event detection tasks \cite{hershey2017cnn}. Such models are typically trained over a mel-spectrogram representation of a waveform that is then treated as an image. Recognizing vocal burst categories (e.g., laughter, sighs) could be seen as a special-case of the sound event detection problem, and a small set of vocal burst categories are contained in AudioSet's ``human voice'' ontology branch \cite{gemmeke2017audio}. While these models may be suitable for vocal burst recognition, it is unclear whether they are suitable for recognizing the emotional nuance in vocal bursts.

We also explore features extracted using large pre-trained self-supervised Conformers \cite{gulati2020conformer} to better understand the performance headroom available on this dataset. In particular, we use a 600M+ parameter Conformer model trained on a 900K hour dataset of ``speech-heavy'' YouTube videos (YT-U) \cite{zhang2021bigssl}. This Conformer has been shown to achieve SoTA on a number of speech paralinguistic tasks including speech emotion recognition and mask detection \cite{shor2022universal}.

Finally, we explore the relationship between the three tasks of perceived emotion, age, and country of origin prediction. In the literature there is evidence of cultural modulation and in-group advantage for emotion recognition across multiple channels~\cite{elfenbein2002universality}. However, on \textsc{HUME-VB}, we observed little evidence for the relationship between the three tasks in the context of the two model families we studied. Our best competition models predict each task individually with no shared parameters.

\section{Model Architectures}

\begin{table*}[t]
\caption{Performance comparison of models trained on \textsc{HUME-VB}.}
\label{tab:mtl}
\vskip 0.15in
\begin{center}
\begin{small}
\begin{sc}
\begin{tabular}{llllll}
\toprule
Model                    & MTL? & $\uparrow$ Emotions\ (CCC)  & $\uparrow$ Country\ (UAR) & $\downarrow$ Age\ (MAE) & $\uparrow$ Harmonic\ mean \\
\midrule
Naive                 & --- & --- & 0.250  & 3.778 & --- \\
Baseline~\cite{baird2022icml}                 & YES & 0.416          & 0.506        & 4.422     & 0.349      \\
\midrule
% \multicolumn{5}{c}{Binary cross-entropy loss for emotion classification} \\
%\midrule
%ResNet50                 & 0.5419          & 0.4877        & 4.063     & 0.3769        \\
%ResNet34                 & 0.5383 & 0.4690 & 4.007 & 0.3752 \\
% %ResNet18                 & 0.5568    & 0.4763  & 4.181 & 0.3714  \\
% \midrule
% Conformer           & 0.6258          & 0.5841        & 3.661     & 0.4304        \\
% Conformer + FC (128)      & 0.6234          & 0.5859        & 3.886     & 0.4168        \\
% Conformer + LSTM     & 0.5323          & 0.5379        & 4.140      & 0.3808        \\
% Conformer  + NetVLAD & 0.6109          & 0.5609        & 4.180      & 0.3947        \\
% Conformer + Autopool & 0.6144          &0.5939        & 3.820      & 0.4207       \\
% \midrule
% \multicolumn{5}{c}{Mean CCC loss for emotion classification} \\
% \midrule
ResNet50                 & YES & 0.569          & 0.513        & 4.093     & 0.385        \\
ResNet50                 & NO & 0.620 & 0.540 & 3.818 & 0.412        \\
ResNet34                 & YES & 0.587 & 0.483 & 4.140 & 0.379 \\
ResNet34                & NO & 0.645 & 0.528 & 3.799 & 0.414 \\
ResNet18             & YES & 0.583    & 0.495  & 4.220 & 0.377  \\
ResNet18             & NO & 0.642	& 0.539 & 3.806 & 0.416  \\
\midrule
Conformer          & YES &    0.647     & 0.572      &   3.780   &        0.424 \\
Conformer          & NO &    0.648     & 0.596      &   3.722   &        0.432 \\
Conformer + FC (128)     & YES & 0.647         & 0.586        & 3.874     & 0.421        \\
Conformer + LSTM      & YES & 0.601         & 0.536        & 4.121     & 0.392        \\
Conformer + NetVLAD     & YES & 0.640         & 0.594        & 3.910     & 0.419        \\
Conformer + AutoPool     & YES & 0.652         & 0.587        & 3.954     & 0.417        \\
\bottomrule
\end{tabular}
\end{sc}
\end{small}
\end{center}
\vskip -0.1in
\end{table*}
\subsection{ResNet models}

We downsample audio waveforms to 16kHz and normalize the signal between $-1$ and $1$ by re-scaling using the maximum and minimum values. On these normalized waveforms, we compute log mel-spectrograms using an FFT frame window of 64 ms and a frame step of 24 ms. We use 128 mel filters spaced between 0 kHz and 8 kHz.

We use each of ResNet18, ResNet34, and ResNet50 architectures as our shared cross-task model trunk with hard parameter sharing. We average pool the final convolutional layer of the ResNet, on top of which we apply dropout and dense layers projecting to each individual task. For the emotional intensity prediction task, we use a linear activation with a negative mean concordance correlation coefficient (\texttt{meanCCC}) loss. For the age classification task, we use a linear activation with a Mean-Squared-Error loss. Finally, for the country of origin prediction task, we use a softmax activation and a categorical cross-entropy loss.

When training all three tasks simultaneously, the emotion task under-trained in comparison to the other two tasks. So, we experimented with a task-specific weighted loss. We found the most success with a weight of $5.0$ for the emotion task, and a weight of $0.05$ for both the age and country of origin tasks. We trained using Adam \cite{kingma2014adam} with a learning rate of $3e{-}4$ and epsilon of $1e{-}08$ with batch sizes of $64$. We found that when training using the \texttt{meanCCC} objective, larger batch sizes stabilized the loss.

\subsection{Conformer-based models}

We preprocess the audio by downsampling to 16kHz and normalizing between $\pm32768.0$. The Conformer model takes as input 80-bin log mel spectrogram features that are first processed by a 3-layer 1D convolutional feature encoder, and subsequently processed by two convolutional strides of two that creates a vector time series that has been downsampled by a factor of 4. On this vector time series a sequence of 24 Conformer layers are stacked to produce an encoding for each time point. The Conformer architecture is pre-trained using the Wav2Vec 2.0 contrastive loss \cite{baevski2020wav2vec} on the ``speech-heavy'' YT-U dataset \cite{zhang2021bigssl}.

In all our experiments, we freeze the base Conformer model developed in~\cite{shor2022universal} and use the 1024-dimensional embeddings from the 12-th layer of the Conformer as input features. The 12-th layer of the model was shown in previous experiments to be a powerful ``universal embedding'' for a variety of downstream paralinguistic tasks~\cite{shor2022universal}. We extract one embedding for approximately $2.5$s non-overlapping window of the audio.

We trained five multitask learning (MTL) models, both simple fully connected models on the aggregated 1024-dimensional (Conformer) features as well as sequence models: (1) \texttt{Conformer}: the features are averaged across the entire audio clip and three dense layers projecting to individual classification/regression tasks (i.e., \textit{task layers}),
(2) \texttt{Conformer+FC}: A single \textit{shared} fully connected (FC) layer of 128 units on top of the averaged feature, followed by the task layers.
(2) \texttt{Conformer+LSTM}: A single LSTM layer of 128 units on top of the features, followed by the task layers, 
(3) \texttt{Conformer+NetVlad}: A NetVlad layer~\cite{arandjelovic2016netvlad} of 5 clusters followed by the task layers, and 
(4) \texttt{Conformer+Autopool}: An adaptive pooling layer~\cite{mcfee2018adaptive} which learns a set of weights before aggregating the features, followed by the task layers.
Besides LSTM, we chose NetVlad and Autopool as they have shown competitive performance, and fewer trainable parameters than that of LSTMs.
We follow the same activation functions and losses as the ResNet models. We trained using Adam \cite{kingma2014adam} with a learning rate of $3e{-}4$, epsilon of $1e{-}08$ and a batch size of $128$.

%\begin{table}[t]
%\caption{Performance comparison of single-task models on \textsc{HUME-VB}.}
%\label{tab:single-task}
%\vskip 0.15in
%\begin{center}
%\begin{small}
%\begin{sc}
%\scalebox{0.78}{
%\begin{tabular}{llll|l}
%\toprule
%Task / model  & Conformer             & ResNet34 & ResNet18             & Baseline %    \\
%\midrule
%Emotions (CCC) &       \textbf{0.6481 }                    & 0.6453  & 0.6416 & %0.4160\\
%Country (UAR) & \textbf{0.5955} &        0.5276 & 0.5394 & 0.5060                  %  \\
%Age (MAE)     & \textbf{3.7220}  &       3.7989 & 3.8055 & 4.4220                \\
%\bottomrule
%\end{tabular}}
%\end{sc}
%\end{small}
%\end{center}
%\vskip -0.1in
%\end{table}

\section{Results}

The results from various MTL models are summarized in Table~\ref{tab:mtl}.
All results are reported on the competition defined validation split.
Per~\cite{baird2022icml}, we use \texttt{meanCCC}, unweighted average recall (\texttt{UAR}) and mean absolute error (\texttt{MAE}) for evaluating the performance of emotion intensity, country and age prediction respectively. We also report the harmonic mean of the three scores.
We refer the reader to~\cite{baird2022icml} for details on the performance metrics.

The first row reports the naive baseline for each task, computed as follows:
For country prediction, the majority class, ``United States'' was used.
The median age of speakers in training set ($26.0$ years) was used for age prediction.
Notably, for the age classification task, using the median age of the speakers in training split outperformed the \textsc{ExVo} competition baseline~\cite{baird2022icml}, as well as all but one of our proposed MTL models. This is likely due to the narrow range of the age values in the validation set. That is, about two-thirds of the speakers in the validation split were $26\pm4$ years old.

\subsection{Exploring the relationship between tasks}
\begin{figure}[ht]
\vskip 0.2in
\begin{center}
\centerline{\scalebox{0.35}{\includegraphics[]{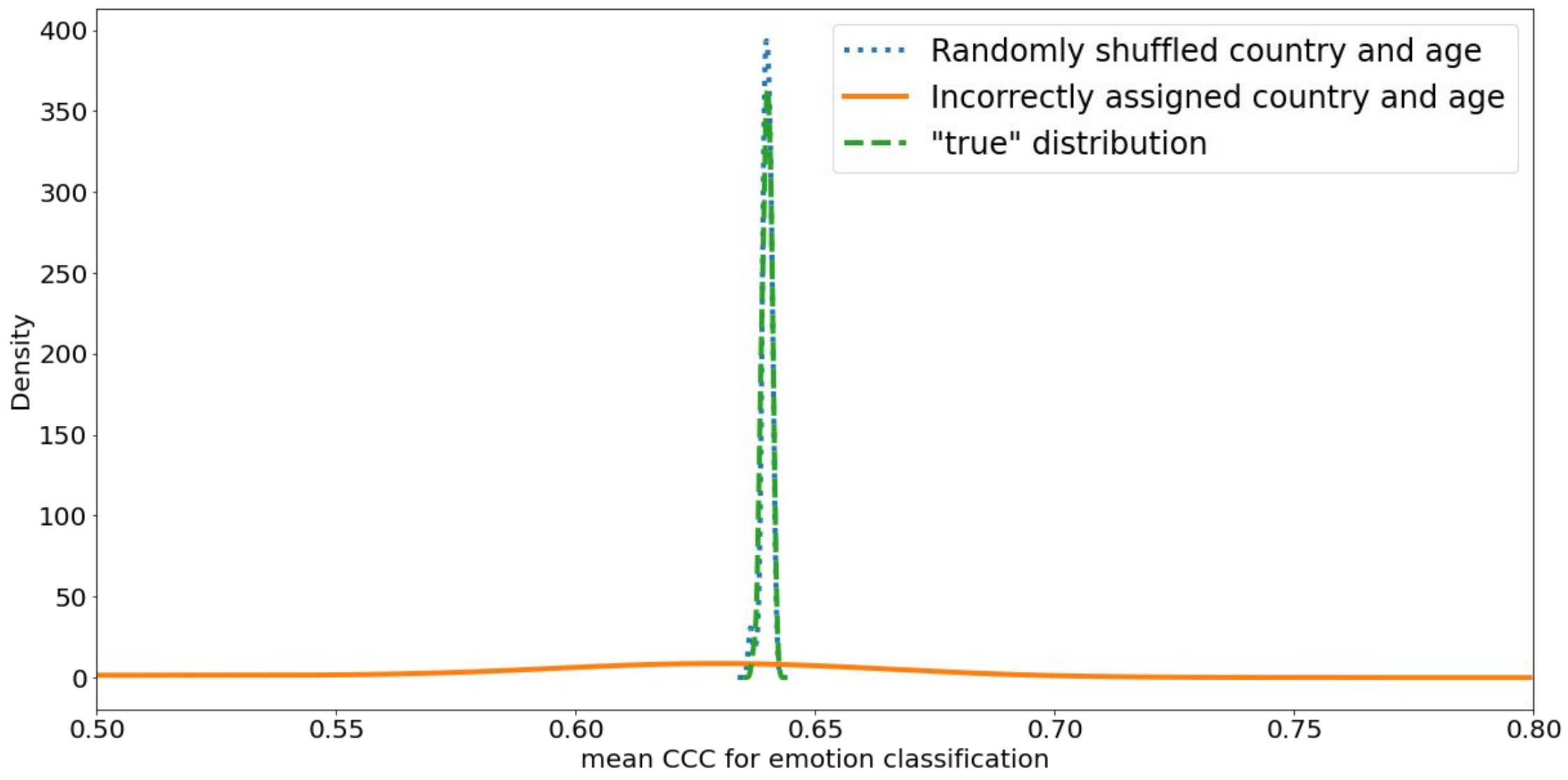}}}
\caption{Gaussian kernel density estimates of emotion classification \texttt{meanCCC} scores across $50$ trials for permutation testing. Country and age were concatenated to the Conformer features. }
\label{fig:multi-input-permutation}
\end{center}
\vskip -0.2in
\end{figure}

In addition to MTL, for many models we explored training each task individually (MTL ``No'' in Table~\ref{tab:mtl}). We did not observe a boost in performance with MTL. In fact, when predicting country of origin, the MTL models performed poorer than their single-task counterparts (e.g., $0.596$ vs $0.583$ for Conformer-based, $0.528$ vs $0.483$ for ResNet34). This suggests that the traditional benefits of MTL, like generalization via regularization and feature sharing for representational efficiency, do not seem to be present for the three tasks considered~\cite{ruder2017overview} within the context of \textsc{HUME-VB}.

In order to better understand the relationship between the three tasks, we experiment with using the labels from two of the three tasks (country of origin and age) as \textit{inputs} to our models. We use permutation testing to compare the performance of a model when given random task labels as input versus when given true task labels as input. Intuitively, if the country and age tasks were useful for perceived emotional intensity prediction, we would expect to see a higher \texttt{meanCCC} score for the true-label model vs the random-label model.
\newline
Concretely, first, we shuffled the country and age values (with different random seeds for each variable) across all samples in the training set while leaving the emotion intensity values unaltered. We then trained a Conformer-based model on this shuffled data with the country and age variables concatenated to the input Conformer features (i.e., total 1049 dimensional input). We repeat this process 50 times to generate a ``random distribution'' of emotional intensity \texttt{meanCCC} scores from the unaltered validation set.
We refer to these models as ``randomly shuffled" (blue dotted line in Fig.~\ref{fig:multi-input-permutation}).

Although shuffling the country and age variables across the entire train split maintains the overall distribution, on average 28\% and 6\% of the samples receive a correct label assignment for country and age respectively after shuffling. In order to further probe to what extent country and age information help improve emotion classification prediction, we randomize the country and age assignment such that they do not follow the original distributions in the train split by assigning incorrect country and age labels for all samples. Like before, we train Conformer-based models on the concatenated inputs, and we repeat this 50 times to generate a distribution of \texttt{meanCCC} scores which we refer to ``incorrectly assigned'' (orange line in Fig.~\ref{fig:multi-input-permutation})

In order to generate a distribution of the \texttt{meanCCC} scores when using the true labels, we repeat the experiment in Table~\ref{tab:mtl} 50 times without altering the corresponding country, age, or emotion labels. Due to SGD and the mini-batch operations, the multiple runs are not deterministic, and thus only produce an approximate ``true distribution'' of \texttt{meanCCC} scores, shown in green dashed line in Fig.~\ref{fig:multi-input-permutation}.

We conducted two-sample t-tests to assess if the performance in the three settings described here was different.
We did not observe a significant difference in emotion classification performance for the true distribution vs. randomly shuffled country and age ($t=0.14,p=0.89$). However we did observe that when the distribution of country and age were changed through incorrect assignment, the unaltered models performed better ($t=4.7,p<0.01$). This analysis suggests that the influence of country and age variables on emotion classification is not substantial. This may explain why we see minimal improvement in performance in the MTL models.

% In the permutation experiments thus far, we are only able to assess if the auxiliary learning significantly benefited the emotional expression classification task within the constraints of the model complexity, model architecture etc. In order to further test our hypothesis more directly, we repeat the experiments described above for the sole task emotional expression classification task but with randomized country and age information concatenated to the input features. Thus the input to these models is a 1029-dimensional vector (1024-dim conformer features, 4-dim one-hot encoded country and age). The results for this permutation testing is shown in Table~\ref{fig:multi-input-permutation}.

\subsection{Conformer-based results}

Our Conformer-based models seem to outperform our ResNet models on all tasks, most notably on the country of origin task ($0.5955$ vs $0.5394$ when trained single-task). However, given the success of large self-supervised representations for downstream acoustic tasks~\cite{hsu2021hubert, shor2022universal, baevski2020wav2vec}, we were surprised to discover that, on the emotional intensity task, our best pre-trained Conformer-based models only marginally outperformed ($0.648$ vs $0.645$) relatively smaller models like ResNet34 that were trained directly on \textsc{HUME-VB} without any pre-training. While this highlights the suitability of the pre-trained Conformer representations for vocal burst modeling, we expected a larger performance gap; especially considering the amount of data used for pre-training Conformers. We have a few hypotheses for this discrepancy.

First, the downstream speech emotion recognition tasks on which the Conformer representations demonstrated SoTA performance---like IEMOCAP \cite{busso2008iemocap} and CREMA-D \cite{cao2014crema}---were largely speech-centric tasks and do not contain many vocal bursts. It is possible that speech emotion recognition based on prosody may not generalize well to the types of emotional expression found in vocal bursts.

Second, the Conformer model was pre-trained on a ``speech-heavy'' dataset that may not contain vocal bursts like those in \textsc{HUME-VB}. In addition, the manner in which these Conformers were pre-trained may not be optimized for modeling the emotional nuances of vocal bursts. If this is the case, the use of pre-trained Conformers to model vocal bursts remains an open area for study.

Finally, it is possible that the Conformer model representations are significantly more powerful at modeling vocal bursts, but \textsc{HUME-VB} is poor at differentiating the performance of these models. For instance, maybe the dataset is noisy in that the emotion intensity labels are not sufficiently explained by the vocalizations themselves, leading to a noise ceiling (e.g., \citealt{lage2019methods}) in possible model performance.

\subsection{Competition Submission}

Our \textsc{ExVo-MultiTask} submission consists of an ensemble of two single-task ResNet34 models, one trained directly on the emotion task and the other, on the country of origin task. We found in our experiments that ResNet34 was the best performing ResNet model on both tasks. Surprisingly, for age, we opted to use the median of $26.0$, as we were unable to reliably outperform with our trained models. We created our own train-validation split by moving vocalizations from 250 random speakers in the \textsc{HUME-VB} validation set to the train split. Our analysis showed that the overall country, age, and emotional intensity distributions remained largely the same in both train and validation after this change. Our submission achieved a harmonic mean of $0.406$ per the MTL competition metric: a 21.24\% increase over the baseline.

\section{Caveats}

Machine perception models of apparent emotional expression, including in vocal bursts, remain an open area of investigation where further research is needed. The models in this work do not aim to infer the internal emotional state of individuals, but rather model proxies in vocalizations. Research in this work was performed expressly to better understand the connections between emotional expression, age, and country of origin, not for any applications; modeling of age and country was solely limited to this purpose and we underscore that a multitask framing with these categories may actually perpetuate bias in some settings. Finally, we emphasize that the observations in this work were limited to the \textsc{HUME-VB} dataset provided in the competition and may not necessarily apply to other vocal burst datasets. This work underwent review for alignment with Google’s AI Principles (https://ai.google/principles/).

\section{Conclusion}

In this work, we demonstrated that ResNet models are competitive in predicting the perceived emotional expression of vocal bursts and other subject attributes. An ensemble of ResNet34 models trained on emotion intensity and country of origin comprised our main submission to \textsc{ExVo-MultiTask}.

We further demonstrated that pre-trained Conformers that have been explored previously for detecting paralinguistic phenomena also perform quite well on all three tasks. However, when predicting perceived emotional intensity, they performed only marginally better than significantly smaller models trained directly on the competition data.

Finally, we explored the relationship between the three competition tasks, and found that models trained on the emotion task with randomly permuted country and age as input features appear to perform approximately the same as models trained with the true country and age as input features. This suggests that there is surprisingly little interaction between the three tasks in \textsc{HUME-VB}. Understanding who the raters were for labeling perceived emotion and their relationship with the stimuli may help disentangle these findings further.
\section{Acknowledgements}
We would like to thank Joel Shor for his guidance in using the conformer model explored in this work.
\newpage
\bibliography{example_paper}
\bibliographystyle{icml2022}

\end{document}